\documentclass[british,10pt]{iopart}
\usepackage{mathptmx}
\usepackage[T1]{fontenc}
\usepackage[utf8]{inputenc}
\usepackage{color}
\usepackage{babel}
\usepackage{amstext}
\usepackage{amssymb}
\usepackage{graphicx}

\makeatletter

\providecommand{\tabularnewline}{\\}

\usepackage{iopams}
\usepackage{setstack}

\makeatother

\begin{document}
\paper{Can we rely on hybrid-DFT energies of solid-state problems with local-DFT
geometries?}
\author{J. D. Gouveia and J. Coutinho}
\address{Department of Physics and I3N, University of Aveiro, 3810-193 Aveiro,
Portugal}
\ead{$^{*}$gouveia@ua.pt}
\begin{abstract}
Hybrid functionals often improve considerably the accuracy of density-functional
calculations, in particular of quantities resulting from the band
structure. In plane-wave (PW) calculations this benefit comes at the
cost of an increase in computation time by several orders of magnitude.
For this reason, large-scale problems addressed within the PW formalism
have to rely on pre-relaxed atomistic geometries, obtained with cheaper
local or semi-local exchange-correlation functionals. We investigate
how suitable these geometries are when plugged into single-point hybrid-DFT
calculations. Based on several case studies, we find two important
sources of error originating from (i) bond strain and (ii) over-mixing
between defect and crystalline states. The first arises from the mismatch
between the pre-relaxed geometry and that obtained after a subsequent
hybrid-DFT-level relaxation. The second occurs when defect states
edging an underestimated band gap artificially mix with crystalline
states, affecting the local bonding character of the defect, and therefore
leading the spurious hybrid-DFT energies. Due to cancelation effects,
the lingering strain contributes little ($\lesssim10$~meV) to the
error bar of quantities based on energy differences of pre-relaxed
structures. The error from state over-mixing does not benefit from
cancelation effects and has to be monitored with caution. \emph{Published
in Electronic Structure }\textbf{\emph{1}}\emph{, 015008 (2019)}.
\texttt{\textbf{\textcolor{blue}{https://doi.org/10.1088/2516-1075/aafc4b}}}
\end{abstract}
\submitto{Electronic Structure}

\maketitle

\section{Introduction}

The use of hybrid density functionals \cite{Becke1993}, which include
a portion of Hartree-Fock exchange (HF-exchange), has proven to be
an invaluable approach to density functional theory (DFT). Within
the solid-state community, the one proposed by Heyd, Scuseria and
Ernzerhof \cite{Heyd2003,Krukau2006} (HSE) has become a rather popular
choice, particularly for diamagnetic and low-spin systems. HSE-like
functionals screen long-ranged and computationally demanding HF-exchange
interactions, essentially by introducing a fraction of HF-exchange
solely within the short-range part of the functional. Long-range contributions
are simply accounted for with a local or semi-local functional, often
the one prescribed by Perdew, Burke and Ernzerhof (PBE) to the generalised
gradient approximation \cite{Perdew1996}.

Besides being able to account for HF-exchange interactions in metals
\cite{Ashcroft1976}, screened hybrid density functionals excel at
describing accurately the band structure of finite-gap crystals \cite{Heyd2005,Bailey2010,Bernasconi2011,Garza2016},
a virtue with huge importance for the study of defects (including
impurities, dislocations and surfaces) with states in the forbidden
gap.

While the calculation of the non-local Fock integral is rather efficient
when using a Gaussian basis (see for instance Ref.~\cite{Bush2011}),
this is not the case for plane waves (PW), which are a popular and
natural choice, particularly among the solid-state community. Plane
wave hybrid-DFT calculations can take thousands of times longer than
using (semi-)local functionals. For instance, we found that, using
the same number of CPUs, one PBE single-point calculation of the oxygen
vacancy in MgO took 28 seconds, against over 9 hours after changing
the exchange-correlation treatment to HSE (around 1200 times as long).
For this reason, the computation of large-scale problems within PW/hybrid-DFT
has to rely on \emph{pre-relaxed} atomic positions obtained using
a cheaper local or semi-local approach.

Many examples of this approach have been reported in the literature,
including in the study of magnetic materials, two-dimensional materials
and surfaces \cite{Noh2014,Supatutkul2017,Zhachuk2018}, as well as
dopants, impurities and radiation-defects in several materials \cite{Szabo2010,Chanier2013,Trinh2013,Colleoni2016,Coutinho2017}.
Significant differences between fully-relaxed PBE- and pre-relaxed
HSE-level results were obtained. For instance, quantities like the
location of defect transition levels or the height of migration barriers
show considerable discrepancies. This raises the following question:
if the fully-relaxed HSE ground state structures were employed instead,
would the result be the same? Or at least within an acceptable error
bar?

In the 1980s, many self-consistent local density calculations were
carried out assuming fixed structures obtained on the basis of known
bond lengths and bond strengths (see for instance Ref.~\cite{Walle1989}).
Despite being less refined, this approximation is similar to using
DFT/pseudopotential pre-relaxed geometries for the calculation of
Mössbauer parameters using all-electron full-potential methods \cite{Wright2016}.
This approach has also been used to obtain the formation energy and
electronic structure of defects using Hedin's $GW$ method \cite{Rinke2009,Avakyan2018}
or diffusion quantum Monte Carlo \cite{Flores2018}. Again, the above
practice raises the question: can we actually rely on energies of
state-of-the-art calculations that employ local-DFT geometries?

\begin{figure}
\begin{centering}
\includegraphics[width=8cm]{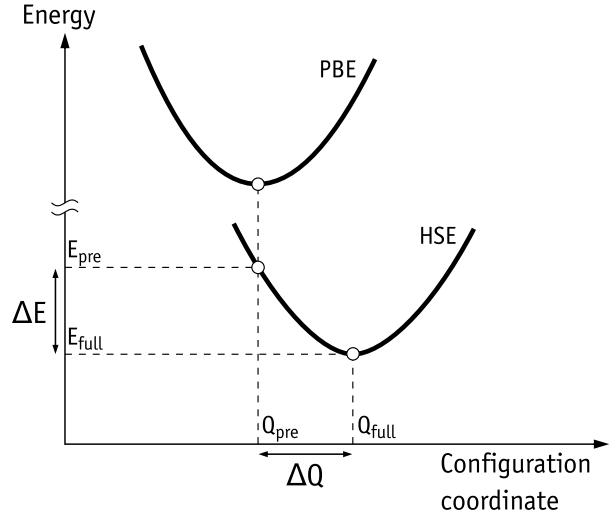}
\par\end{centering}
\caption{\label{fig1}Schematic representation of the relaxation method followed
in this work. The structure is pre-relaxed within PBE-level, down
to the $Q_{\mathrm{pre}}$ structure (upper curve). A subsequent relaxation
using HSE starts with total energy $E_{\mathrm{pre}}$, and yields
the fully-relaxed HSE ground state configuration $Q_{\mathrm{full}}$
with energy $E_{\mathrm{full}}$. $\Delta E$ is the HSE relaxation
energy, and quantifies the error of the pre-relaxed calculation.}
\end{figure}

Let us look at this problem with the help of Figure~\ref{fig1},
where potential energy curves corresponding to sequential PBE- and
HSE-level relaxations are illustrated. A starting (guessed) structure
is subjected to a first relaxation step using the PBE functional.
This provides the pre-relaxed minimum-energy geometry, $Q_{\mathrm{pre}}$,
which is plugged into a single-point HSE calculation to provide the
pre-relaxed state $(Q_{\mathrm{pre}},E_{\mathrm{pre}})$. A subsequent
relaxation step also using HSE drives the system to the final ground
state configuration $(Q_{\mathrm{full}},E_{\mathrm{full}})$. The
use of the pre-relaxed $(Q_{\mathrm{pre}},E_{\mathrm{pre}})$ instead
of the fully-relaxed $(Q_{\mathrm{full}},E_{\mathrm{full}})$ state
is rather tempting. Unfortunately the underlying effects governing
the error $\Delta E=E_{\mathrm{full}}-E_{\mathrm{pre}}$ are not obvious.

Below we find some answers based on the magnitude of spurious strain
fields and resonances between defect- and crystalline-related states,
due to the change of functional.

We address this issue by assessing the accuracy of HSE single-point
calculations performed on pre-relaxed structures. These are compared
to corresponding HSE fully-relaxed calculations. Four case studies
are considered, namely the oxygen vacancy in magnesium oxide, the
oxygen interstitial in silicon, the Si(001) surface, and the carbon
interstitial in silicon carbide. In each case, we calculate $\Delta E$
values, the displacement of the most relevant atoms (closest to the
defect), as well as selected observables. In Section~\ref{sec:method}
we disclose the technical details of the calculations. In Section~\ref{sec:results}
we present and discuss our results. Finally, we draw the conclusions
in Section~\ref{sec:conclusions}.

\section{Method details\label{sec:method}}

First-principles calculations were performed within hybrid and semi-local
density functional theory using the Vienna Ab-initio Simulation Package
(VASP) \cite{Kresse1993,Kresse1994,Kresse1996,Kresse1996a}. The HF-exchange
mixing fraction and screening parameter for the HSE functional were
$a=1/4$ and $\omega=0.2$~Å$^{-1}$, respectively (resulting in
the commonly named HSE06 functional) \cite{Krukau2006}. The projector-augmented
wave method was employed to avoid explicit treatment of core electrons
in the Kohn-Sham equations \cite{Blochl1994}. These were solved self-consistently
using the PW formalism, until the total energy between two consecutive
steps differed by less than $10^{-7}$~eV. The choice of energy cutoff
values for PW, as well as the density of Brillouin-zone (BZ) sampling
meshes, essentially depend on the problem at hand (chemical species,
material of interest and supercell sizes). They were chosen after
convergence tests, ensuring that absolute (relative) energies were
converged within less than 10~meV/atom (less than 1~meV/atom).

HSE lattice parameters were adopted in all four case studies referred
above. Ionic relaxations stopped when the maximum force acting on
every atom became smaller than 0.01~eV/Å. For each case, $(Q_{\mathrm{pre}},E_{\mathrm{pre}})$
and $(Q_{\mathrm{full}},E_{\mathrm{full}})$ pairs were obtained by
means of the above two-step recipe (see Figure~\ref{fig1}), where
$\Delta E=E_{\mathrm{full}}-E_{\mathrm{pre}}$.

For the oxygen vacancy in magnesium oxide (MgO:V$_{\mathrm{O}}$),
we used a 64-atom cubic supercell (minus an oxygen atom). The PW energy
cutoff was converged at $E_{\mathrm{cut}}=500$~eV and a Monkhorst-Pack
(MP) $2\times2\times2$ grid of $\mathbf{k}$-points proved adequate
\cite{Monkhorst1976}. The calculated HSE-level (PBE-level) lattice
parameter was $a_{0}^{\mathrm{HSE}}=4.200$~Å ($a_{0}^{\mathrm{PBE}}=4.255$~Å),
comparing well with the experimental value of $a_{0}^{\mathrm{exp}}=4.216$~Å
\cite{Hirata1977}. The HSE-level (PBE-level) direct band gap was
$E_{\mathrm{g}}^{\mathrm{HSE}}=6.64$~eV ($E_{\mathrm{g}}^{\mathrm{PBE}}=4.73$~eV),
underestimating the observed gap of $E_{\mathrm{g}}^{\mathrm{exp}}=7.8$~eV
by 15\% (40\%) \cite{Whited1969}. The oxygen vacancy (V$_{\mathrm{O}}$)
in MgO is a double donor. We investigated the deviation of the transition
levels obtained from single-point energies (of pre-relaxed structures)
with respect to those from fully-relaxed HSE-level structures. Unwanted
Coulomb interactions between periodic replicas of charged defects
were removed from the total energy according to the method proposed
by Freysoldt, Neugebauer and Van de Walle \cite{Freysoldt2009}.

For the oxygen interstitial (O$_{\mathrm{i}}$) defect in Si, several
structures close to the minimum ground state and to the saddle point
for migration were explored. Supercells with 64 Si atoms (plus one
O atom) were used, and the BZ was sampled using an MP $2\times2\times2$
grid of $\mathbf{k}$-points. The calculated lattice parameter was
$a_{0}^{\mathrm{HSE}}=5.432$~Å ($a_{0}^{\mathrm{PBE}}=5.469$~Å),
also in good agreement with the measured value $a_{0}^{\mathrm{exp}}=5.431$~Å
\cite{Mohr2016}. The Kohn-Sham band gap was estimated as $E_{\mathrm{g}}^{\mathrm{HSE}}=1.15$~eV
($E_{\mathrm{g}}^{\mathrm{PBE}}=0.57$~eV), which is to be compared
with $E_{\mathrm{g}}^{\mathrm{exp}}=1.17$~eV from optical experiments
\cite{Green1990}. The PW energy cutoff was set at $E_{\mathrm{cut}}=400$~eV.

The third case considered was the Si(001) surface, where the energy
difference between symmetric $(2\times1)$ and asymmetric b$(2\times1)$
reconstructions was investigated \cite{Yin1981}. We employed 19-monolayer-thick
symmetric slabs (composed of 38 Si atoms), separated by 11~Å of vacuum
space. An MP $6\times6\times1$ grid of $\mathbf{k}$-points was used
to sample the BZ of both reconstructed surfaces and, like in the previous
case, $E_{\mathrm{cut}}=400$~eV.

Finally, carbon interstitial (C$_{\mathrm{i}}$) defects in 3$C$-SiC
were investigated on cubic supercells with 64 atoms (plus the interstitial
C atom). The energy cutoff was $E_{\mathrm{cut}}=400$~eV and the
BZ sampling was carried out using an MP $4\times4\times4$ grid of
$\mathbf{k}$-points. The calculated lattice parameter and Kohn-Sham
gap were $a_{0}^{\mathrm{HSE}}=4.347$~Å ($a_{0}^{\mathrm{PBE}}=4.380$~Å)
and $E_{\mathrm{g}}^{\mathrm{HSE}}=2.25$~eV ($E_{\mathrm{g}}^{\mathrm{PBE}}=1.33$~eV),
respectively. Again, these compare with $a_{0}^{\mathrm{exp}}=4.360$~Å
and $E_{\mathrm{g}}^{\mathrm{exp}}=2.42$~eV from experiments, as
expected \cite{Bimber1981}. We investigated two structures of the
C$_{\mathrm{i}}$ defect reported in the literature, namely a tilted-$\langle001\rangle$
split-interstitial with $C_{1h}$ symmetry \cite{Bockstedte2003}
and an upright-$\langle001\rangle$ split-interstitial with $D_{2d}$
symmetry (or simply $\langle001\rangle$ split-interstitial) \cite{Gali2003}.
The latter comprises a $\langle001\rangle$-aligned C-C dimer at the
carbon site (both C atoms are three-fold coordinated and symmetrically
equivalent), while in the monoclinic structure the C-C bond makes
an angle with the $\langle100\rangle$ axis, ending up in a structure
where one of the C atoms has four-fold coordination, while the other
keeps the three-fold coordination.

\section{Results and discussion\label{sec:results}}

\subsection{Oxygen vacancy in MgO}

The oxygen vacancy in MgO has the same symmetry ($O_{h}$) in all
stable charge states ($q=0,\,+1$ and $+2$). However, the displacement
of its neighbours (from their crystallographic positions) is variable.
Accordingly, whereas Mg$^{2+}$ first neighbours are pushed away from
the vacant site by $d\approx0.085$~Å per ionised electron, O$^{2-}$
next neighbours are attracted to the centre by $d\approx0.035$~Å
for each ionisation. These results are also close to those reported
by Rinke and co-workers \cite{Rinke2012} using the local density
approximation, where outward and inward displacements of about 0.09~Å
and 0.03~Å per ionisation were reported for Mg and O atoms, respectively.

In the top-most section of Table~\ref{tab1}, we report the displacement
($d$) of Mg and O nearest neighbours to V$_{\mathrm{O}}$ from fully-relaxed
geometries with respect to pre-relaxed ones. Results are shown for
neutral, positively and double positively charged defects. Pre-relaxed
and fully-relaxed defect geometries are very similar. The largest
atomic displacement with respect to the pre-relaxed geometry was 0.015~Å,
and that was observed for the Mg first neighbours. Still, after relaxation,
total energies differed by at most $\Delta E\sim-40$~meV from those
of pre-relaxed structures.

\begin{table*}
\caption{\label{tab1}Four data sets related to (1) V$_{\mathrm{O}}$ in MgO
in different charge states, (2) four configurations of O$_{\mathrm{i}}$
in Si, (3) two reconstructions for the Si(001) surface, and (4) two
configurations of C$_{\mathrm{i}}$ in $3C$-SiC. First and second
data rows of each set show displacement magnitudes ($d$) of selected
atoms after full HSE relaxation, relatively to pre-relaxed positions.
The third row of each data set reports the relaxation energy $\Delta E$
or surface relaxation energy $\Delta\sigma=\Delta E/2A$ of pre-relaxed
structures, where $A$ is the surface unit cell area. The fourth and
fifth rows of each data set show HSE results using pre-relaxed and
fully-relaxed geometries. These include transition levels $E(q/q\!+\!1)-E_{\mathrm{c}}$
with respect to the conduction band bottom, relative energies $E-E_{\mathrm{GS}}$
with respect to the ground state, and surface formation energies,
$\sigma$ (see text).}

\centering{}%
\begin{tabular}{cccccc}
\hline 
MgO:V$_{\mathrm{O}}$ & Units & $q=0$ & $q=+1$ & $q=+2$ & \tabularnewline
$d$(Mg) & Å & 0.007 & 0.010 & 0.015 & \tabularnewline
$d$(O) & Å & 0.001 & 0.001 & 0.001 & \tabularnewline
$\Delta E$ & eV & $-0.003$ & $-0.012$ & $-0.038$ & \tabularnewline
$E_{\mathrm{\text{pre}}}(q/q+1)-E_{\mathrm{c}}$ & eV & 2.872 & 4.190 & --- & \tabularnewline
$E_{\mathrm{full}}(q/q+1)-E_{\mathrm{c}}$ & eV & 2.863 & 4.163 & --- & \tabularnewline
\hline 
Si:O$_{\mathrm{i}}$ & Units & $C_{1h}$ & $C_{2}$ & $D_{3d}$ & $C_{2v}$\tabularnewline
$d$(O) & Å & 0.129 & 0.127 & 0.000 & 0.045\tabularnewline
$d$(Si) & Å & 0.012 & 0.012 & 0.014 & 0.027\tabularnewline
$\Delta E$ & eV & $-0.019$ & $-0.018$ & $-0.009$ & $-0.013$\tabularnewline
$E_{\mathrm{pre}}-E_{\mathrm{pre},\mathrm{GS}}$ & eV & 0.000 & 0.002 & 0.001 & 2.758\tabularnewline
$E_{\mathrm{full}}-E_{\mathrm{full},\mathrm{GS}}$ & eV & 0.000 & 0.002 & 0.010 & 2.764\tabularnewline
\hline 
Si(001) & Units & $(2\times1)$ & b$(2\times1)$ &  & \tabularnewline
$d$(Si$_{1}$) & Å & 0.147 & 0.115 &  & \tabularnewline
$d$(Si$_{2}$) & Å & 0.147 & 0.164 &  & \tabularnewline
$\Delta\sigma$ & meV/Å$^{2}$ & $-1.048$ & $-0.856$ &  & \tabularnewline
$\sigma_{\mathrm{pre}}$ & meV/Å$^{2}$ & 97.23 & 95.39 &  & \tabularnewline
$\sigma_{\mathrm{full}}$ & meV/Å$^{2}$ & 96.18 & 94.53 &  & \tabularnewline
\hline 
$3C$-SiC:C$_{\mathrm{i}}$ & Units & $q=0$ & $q=+1$ &  & \tabularnewline
$d$(C$_{1}$) & Å & 0.013 & 0.000 &  & \tabularnewline
$d$(C$_{2}$) & Å & 0.539 & 0.000 &  & \tabularnewline
$\Delta E$ & eV & $-0.101$ & $-0.005$ &  & \tabularnewline
$E_{\mathrm{\text{pre}}}(q/q+1)-E_{\mathrm{c}}$ & eV & 0.367 & --- &  & \tabularnewline
$E_{\mathrm{full}}(q/q+1)-E_{\mathrm{c}}$ & eV & 0.476 & --- &  & \tabularnewline
\hline 
\end{tabular}
\end{table*}

In the above, we inspected the quality of pre-relaxed structures for
the calculation of HSE total energies. The same analysis could be
done for the calculation of defect formation energies,

\begin{equation}
E_{\mathrm{f}}(q)=E(q)-\sum_{i}n_{i}\mu_{i}+q(E_{\mathrm{v}}+E_{\mathrm{F}}).\label{eq:fe}
\end{equation}
This is the amount of energy required to combine $n_{i}$ elements
of species $i$ and form a defective crystal (with computed energy
$E$). Elements are assumed to be available in standard phases with
respective chemical potentials $\mu_{i}$ (see Ref.~\cite{Freysoldt2014}
and references therein for further details). In Eq.~\ref{eq:fe},
$q$ is the charge state of the defect, obtained by exchanging electrons
between an electronic reservoir with chemical potential $E_{\mathrm{v}}+E_{\mathrm{F}}$
(where $E_{\mathrm{F}}$ is the Fermi level with respect to the valence
band top energy, $E_{\mathrm{v}}$) and the highest occupied or unoccupied
states. It is assumed that the calculation of $\mu_{i}$ and $E_{\mathrm{v}}$
in Eq.~\ref{eq:fe} does not involve a pre-relaxation step. Hence,
the error of the calculated formation energies using pre-relaxed structures
is the same as that of total energies, $\Delta E_{\mathrm{f}}\equiv\Delta E$.

We emphasise that $\Delta E\leq0$ , so that the error involving the
energy difference between two pre-relaxed structures benefits from
cancelation effects. An example of such a quantity is a defect transition
level, which by definition, is given by the location of $E_{\mathrm{F}}$
where $E_{\mathrm{f}}(q)=E_{\mathrm{f}}(q+1)$. In the case of a double
donor such as MgO:V$_{\mathrm{O}}$, the $(q+1)$-th donor level (with
respect to the conduction band minimum, $E_{\mathrm{c}}$) is given
by $E_{\mathrm{c}}-E(q/q+1)=[\epsilon_{\mathrm{c}}+E(q+1)]-E(q)$,
where $\epsilon_{\mathrm{c}}$ is the lowest unoccupied state in a
bulk calculation. Hence, for MgO:V$_{\mathrm{O}}$ we obtain $E_{\mathrm{c}}-E(0/+)=2.86$~eV
and $E_{\mathrm{c}}-E(+/\!+\!+)=4.16$~eV using fully-relaxed HSE
energies. These figures would be closer to those obtained from $G_{0}W_{0}$
quasi-particle energies \cite{Rinke2012} should we have accounted
for Frank-Condon relaxation effects and for a better correlation treatment
to improve the gap width. However, these issues are not relevant for
the present analysis. We are interested in assessing the quality of
$E_{\mathrm{c}}-E_{\mathrm{pre}}(q/q+1)$ obtained from pre-relaxed
energies, with respect to the analogous calculation using fully-relaxed
structures, $E_{\mathrm{c}}-E_{\mathrm{full}}(q/q+1)$. The results
are shown in Table~\ref{tab1}. The error bar of the pre-relaxed
results is of the order of few tens of meV, which is quite acceptable
for semiconductors and insulators with a gap width in the eV range.

\subsection{Interstitial oxygen in silicon}

Our second case is a well-established defect in crystalline silicon,
namely Si:O$_{\mathrm{i}}$. In the ground state, O$_{\mathrm{i}}$
adopts a puckered bond-centre configuration \cite{Coutinho2000}.
The potential energy surface for rotation of the O atom around the
Si-Si bond has the shape of a flat `Mexican hat' (with small bumps
in the meV range). Accordingly, the O atom is slightly displaced (0.3~Å)
away from the centre of a $[111]$-aligned Si-Si bond, either along
$[1\bar{1}0]$ or along $[11\bar{2}]$ (resulting in defects with
$C_{2}$ or $C_{1h}$ symmetry, respectively). The perfect bond-centred
structure has $D_{3d}$ symmetry and it is a maximum in the potential
energy landscape.

Long-range diffusion of O$_{\mathrm{i}}$ in Si occurs via sequential
jumps between neighbouring puckered configurations. At the saddle-point,
the structure passes close to a $C_{2v}$-symmetric configuration
(often referred to as `Y-lid'), consisting of a $\langle100\rangle$-aligned
O-Si dimer sharing a Si site, where both O and Si atoms are three-fold
coordinated \cite{Coutinho2000}. The relaxation of this structure
was achieved with help of force symmetrisation.

The results are shown in Table~\ref{tab1}. Concerning HSE-relaxed
energies, we obtained the following relative energies with respect
to the $C_{1h}$ ground state: $E_{\mathrm{full}}-E_{\mathrm{full},\mathrm{GS}}=2$~meV,
10~meV and 2.76~eV for the $C_{2}$, $D_{3d}$ and $C_{2v}$ structures,
respectively, confirming the flatness of the potential around the
bond centre. The energy of the $C_{2v}$ structure is consistent with
analogous results obtained by Binder and Pasquarello \cite{Binder2014},
where it was demonstrated that hybrid-DFT was able account for the
observed 2.53~eV migration barrier of O$_{\mathrm{i}}$ in silicon
\cite{Corbett1964}. This contrasts with local and semi-local functionals
which predict a barrier of about 2~eV \cite{Coutinho2000}.

Unlike MgO:V$_{\mathrm{O}}$, pre-relaxed structures of O$_{\mathrm{i}}$
in Si hold a \emph{lingering strain} when plugged into a HSE calculation.
This effect is unavoidable and stems from the fact that PBE over-estimates
Si-O bond lengths with respect to those from HSE. The result is the
straightening of the Si-O-Si unit after HSE full relaxation, with
the O atom approaching the bond centre site.

Due to the soft nature of the Si-O-Si bending potential, the above
displacements lead to tolerable energy changes. Values of $\Delta E$
in Table~\ref{tab1} show that fully relaxed HSE total energies are
at most a few tens of meV below those obtained from single-point calculations
performed on pre-relaxed structures. Interestingly, and again due
to cancelation effects, the difference between relative energies whether
using fully-relaxed ($E_{\mathrm{full}}-E_{\mathrm{full},\mathrm{GS}}$)
or pre-relaxed ($E_{\mathrm{pre}}-E_{\mathrm{pre},\mathrm{GS}}$)
structures is negligible (<10~meV).

\subsection{Si(001)-$\mathit{(2\times1)}$ surface}

The Si(001) surface reconstruction is known to be made of dimers,
each possessing two unsaturated radicals. Experiments show that charge
transfer between these radicals leads to dimer buckling \cite{Yin1981},
where one of the atoms becomes protruded, while the other drops into
the surface. The smallest surface unit cell which is able to capture
this effect contains a single dimer. We can either calculate a symmetric
Si(001)-$(2\times1)$ reconstruction, where the dimerised Si atoms
have the same height (no buckling), or the buckled Si(001)-b$(2\times1)$.

Table~\ref{tab1} shows that, for both reconstructions, atom displacements
of pre-relaxed slabs are large. Large displacements were also found
for subsurface Si atoms.

The surface formation energy, $\sigma$, was obtained as $2A\sigma=E-\sum n_{i}\mu_{i}$,
where $A$ is the area of the surface unit cell and the factor of
two accounts for the identical and opposite-facing surfaces on the
slab with total energy $E$. Fully-relaxed surface formation energies
are about $|\Delta\sigma|\sim1$~meV/Å$^{2}$ lower than pre-relaxed
analogues, a value which is regarded too large to make pre-relaxed
energies reliable for absolute formation energy calculations.

Such strong relaxation energies and displacements arise from a mismatch
between PBE- and HSE-level lattice parameters of about 0.7\%, which
leads to a corresponding contraction of the whole slab thickness upon
full HSE-relaxation.

Despite the above, relative energies benefit from cancelation effects.
Table~\ref{tab1} shows that Si(001)-b$(2\times1)$ is more stable
than Si(001)-$(2\times1)$ by 1.84~meV/Å and 1.65~meV/Å when using
pre-relaxed and fully-relaxed energies ($\sigma_{\mathrm{pre}}$ and
$\sigma_{\mathrm{full}}$), respectively.

\subsection{Carbon self-interstitial in in 3C-SiC}

The stable configuration of the carbon self-interstitial in $3C$-SiC
has been investigated concurrently by different groups, with different
structures being proposed for the neutral charge state. Gali \emph{et~al.}
\cite{Gali2003} reported a spin-1 split-interstitial with $D_{2d}$
symmetry, being made of a $\langle001\rangle$-aligned C-C dimer on
a carbon site. The electronic structure consisted on a semi-occupied
doublet state arising from orthogonal $\pi$-like orbitals on both
three-fold coordinated carbon atoms.

Conversely, Bockstedte \emph{et~al.} \cite{Bockstedte2003} found
a diamagnetic monoclinic structure ($C_{1h}$ symmetry), where the
C$_{1}$-C$_{2}$ dimer was tilted towards the $\langle110\rangle$
direction, resulting in three-fold and four-fold coordinated atoms,
respectively. We note that, while the paramagnetic state was obtained
from a post-corrected hybrid-DFT calculation, the diamagnetic and
low-symmetry state was found within PBE.

Our calculations confirm the above conflicting results. PBE- and HSE-relaxations
lead to C$_{\mathrm{i}}(C_{1h})$ and C$_{\mathrm{i}}(D_{2d})$ ground
state structures with total spin $S=0$ and $S=1$, respectively.
Therefore, after obtaining the pre-relaxed structure ($C_{1h}$),
a subsequent HSE calculation displaces the four-fold coordinated C$_{2}$
atom by more than 0.5~Å, raising the symmetry to $D_{2d}$ and flipping
the spin to $S=1$ (see Table~\ref{tab1}). The difference between
pre-relaxed and fully-relaxed HSE energies is as much as $\Delta E=-0.1$~eV.
Obviously this poses a major problem to the use of the pre-relaxed
structures. We will return to this issue below.

In the positive charge state, the C$_{\mathrm{i}}$ defect suffers
a weak Jahn-Teller distortion involving a dynamic overlap of the $\pi$-orbitals
\cite{Bockstedte2003,Gali2003}, and the symmetry is lowered from
$D_{2d}$ to $D_{2}$. Pre-relaxed and fully-relaxed structures show
essentially the same geometry (as shown by the small atomic displacements
in the bottom data set of Table~\ref{tab1}), and the relaxation
energy is only $\Delta E=-5$~meV.

Due to the large relaxation energy obtained for the neutral charge
state, the calculation of the donor level using pre-relaxed energies
does not profit from error cancelation effects. The result differs
from the fully-relaxed donor level by about 0.1~eV (ten times larger
than MgO:V$_{\mathrm{O}}$).

\begin{figure}
\begin{centering}
\includegraphics[width=8.5cm]{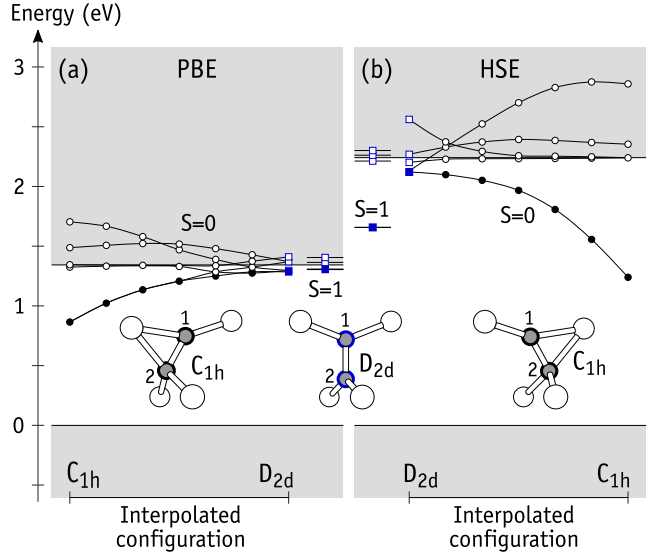}
\par\end{centering}
\caption{\label{fig2}Kohn-Sham eigenvalues (symbols) of linearly interpolated
C$_{\mathrm{i}}$ defects in 512-atom SiC supercells at $\mathbf{k}=\Gamma$.
(a) and (b) represent PBE- and HSE-level results, respectively. Only
the highest occupied (closed symbols) and the next four unoccupied
levels (open symbols) are shown. Circles and squares represent eigenvalues
from $C_{1h}$- and $D_{2d}$-symmetric structures, respectively.
Data points connected by Bézier lines correspond to interpolated diamagnetic
($S=0$) states, while both data sets in the middle correspond to
paramagnetic ($S=1$) states. Valence band and conduction band states
of the crystal are represented by shaded regions.}
\end{figure}

We investigated the origin of the structure/spin disparity between
the above PBE and hybrid-DFT calculations. In order to rule out dispersion
effects due to the finite size of the supercell, we also carried out
216-atom and 512-atom supercell calculations (plus one C atom), using
$2\times2\times2$ and $1\times1\times1$ ($\Gamma$-point) $\mathbf{k}$-point
grids for BZ sampling. Full HSE relaxation was not possible for these
cells. Instead, the $D_{2d}$ structure was subjected to a symmetry-constrained
PBE relaxation, followed by an HSE single-point calculation. Irrespectively
of the supercell size, the HSE energy of the pre-relaxed C$_{\mathrm{i}}(D_{2d},S=1)$
state was lower than that of C$_{\mathrm{i}}(C_{1h},S=0)$ by 0.1~eV.

We went on and inspected the Kohn-Sham levels of seven neutral diamagnetic
structures, obtained after linear interpolation between C$_{\mathrm{i}}(C_{1h})$
and C$_{\mathrm{i}}(D_{2d})$. Analysis of defect levels is more conveniently
done at the $\mathbf{k}=\Gamma$ point, where the wave-functions are
real. However, in order to preserve the BZ-sampling quality, large
cubic supercells with 512 atoms (plus one carbon atom) were employed
for PBE-relaxations and respective HSE single-point calculations.

The results are shown in Figure~\ref{fig2} by data points connected
by Bézier curves for better perception. Circles and squares apply
to $C_{1h}$ and $D_{2d}$ defect symmetries, respectively. Also for
the sake of clarity, only the highest occupied (closed symbols) and
the next four unoccupied levels (open symbols) are shown. Left- and
right-hand sides of the figure refer to results obtained with PBE
and HSE exchange-correlation functionals, respectively. The valence
band top was aligned on both sides at the origin of the energy scale.

The paramagnetic state C$_{\mathrm{i}}(D_{2d},S=1)$ was also investigated.
Its electronic structure is shown in the middle region of the figure
(for both PBE and HSE functionals).

If we ignore the difference in the gap width, the electronic structure
of C$_{\mathrm{i}}(C_{1h},S=0)$ is rather similar whether it is calculated
using PBE or HSE (left and right edges of Figure~\ref{fig2}, respectively).
The defect is a symmetric singlet state ($A$ within $C_{1h}$ point
group), with a fully occupied p-like deep level localised on C$_{1}$
and C$_{2}$ atoms (see inset of \ref{fig2}). As the structure evolves
to $D_{2d}$, the $A$-state mixes with the conduction band states
and the defect becomes a shallow donor. Note that C$_{\mathrm{i}}(D_{2d},S=0)$
is not stable -- within PBE-level, atomic relaxation drives the geometry
back to the $C_{1h}$ structure, while within HSE the ground state
is C$_{\mathrm{i}}(D_{2d},S=1)$.

Hence, comparing PBE and HSE results in Figure~\ref{fig2}, we readily
conclude that within PBE the exchange interactions are underestimated
due to the strong resonance between the doubly degenerate $E$-level
and the low-lying conduction band states. For that reason, the PBE
functional fails to describe the ground state structure of neutral
C$_{\mathrm{i}}$ in $3C$-SiC, and that undermines any pre-relaxed
HSE calculation.

\section{Conclusions\label{sec:conclusions}}

We investigated the suitability of atomistic geometries, particularly
of defects in semiconductors and insulators, obtained within a semi-local
DFT method (referred to as \emph{pre-relaxed} structures), to be used
on single-point hybrid-DFT calculations. To that end, four distinct
case studies were investigated, namely the oxygen vacancy in magnesium
oxide, the oxygen interstitial in silicon, the Si(001) surface and
the carbon self-interstitial in cubic silicon carbide.

We found at least two important sources of error that should be monitored.
The first is the presence of lingering strain within the pre-relaxed
structure, which will be released should a full hybrid-DFT calculation
be performed. The relaxation energy, $\Delta E$, arises from slight
differences in equilibrium bond lengths as obtained from semi-local
and hybrid-DFT methods. It is interpreted as the error bar of single-point
HSE energies based on pre-relaxed structures, including of formation
energies.

The magnitude of $\Delta E$ is system-dependent. For localised point
defects, the effect was estimated to be in the range of a few tens
of meV. This is acceptable for the calculation of most defect-related
observables, including formation energies. Extended defects, on the
other hand, are expected to be affected by larger errors. The surface
formation energy of two Si(001) pre-relaxed reconstructions were calculated
with a discrepancy of $\Delta\sigma~\sim-1$~meV/Å. This is close
to the difference between the surface formation energies of symmetric
and buckled dimerised Si(001)-$(2\times1)$ reconstructions, and larger
than the usual error bar needed for this type of calculation. The
problem arises from the lattice mismatch generated by the different
functionals, that leads to the accumulation of lingering strain across
a large volume of the pre-relaxed geometry (particularly in bulk-like
regions). The presence of vacuum in the slab allows the strain to
relax, releasing relatively large amounts of energy during a full
hybrid-DFT relaxation.

We also found that calculations based on energy differences of pre-relaxed
structures benefit from error cancelation effects. This is because
$\Delta E$ is always negative. The error bar in this case was about
one order of magnitude lower than that affecting total energies. This
feature applies for instance to transition levels, binding energies,
migration barriers, but also to surface formation energy differences.

The second source of error results from an over-mixing of defect states
with the host bands during the pre-relaxation stage. This effect also
depends strongly on the problem at hand. Local and semi-local functionals
are known to underestimate the width of band gaps of insulators and
semiconductors. This favours resonances involving defect levels edging
conduction band and valence band states. The result is rather similar
to the pseudo-Jahn-Teller effect and, as such, leads to artificial
bond formation and breaking. Obviously, the spurious pre-relaxed geometries
will lead to misleading single-point hybrid-DFT energies. We discuss
the effect in the light of a detailed analysis of the carbon interstitial
in $3C$-SiC. Many examples which are expected to show analogous resonances
have been reported in the literature. These include the negative carbon
vacancy in $4H$-SiC \cite{Trinh2013}, or the cadmium vacancy in
CdTe \cite{Flores2018}.

\ack{}{}

This work is supported by the NATO SPS programme, project number 985215.
JC thanks the Fundação para a Ciência e a Tecnologia (FCT) for support
under project UID/CTM/50025/2013, co-funded by FEDER funds through
the COMPETE 2020 Program. JDG acknowledges the financial support from
the I3N through grant BPD-11(5017/2018).

\section*{References}{}

\bibliographystyle{iopart-num}

\providecommand{\newblock}{}

\end{document}